\documentclass{article}
\usepackage{graphicx} 
\usepackage{natbib}
\usepackage{fullpage}
\usepackage{multicol}
\usepackage{xcolor}
\usepackage{hyperref}
\usepackage{booktabs}
\usepackage{subcaption}

\title{A Mathematical Lens for Teaching Data Science}
\author{Jo Hardin}
\date{\today}

\begin{document}

\maketitle

\section{Introduction}
To many of us, data science might suddenly feel ubiquitous. Flagship programs at UC Berkeley, NYU, MIT, and the University of Michigan were developed in the 2010s.  Since then, \href{https://www.datascienceprograms.org/}{https://www.datascienceprograms.org/} now tracks more than 1,000 different data science programs in the United States. The National Center for Education Statistics reported a recent jump in data science bachelor's degrees awarded, from 84 in 2020 to 897 in 2022, almost 11 times as many degrees in two years \citep{Pierson2023}! The programs are often grown out of mathematics or statistics departments (sometimes computer science departments) by individuals who were trained in the mathematical sciences.  Additionally, there has been rapid growth in data science programs across the K-12 curriculum \citep{NAS2023,israelfishelson2024,ds4everyone}.

These developments — the surge in undergraduate programs, the dramatic increase in bachelor’s degrees awarded, and the expansion of data science into K–12 education — can make it {\em feel} like data science is just math or statistics (or computer science).  But, on the other hand, data science really isn't just math or statistics (or computer science). As individuals who have been trained in the mathematical sciences (us), it is important to understand the ways that a data science curriculum connects directly to a mathematics curriculum and the ways in which data science is quite distinct from curricula in the mathematical sciences. Indeed, the training that students receive within the mathematical sciences is often aligned with the building blocks of data science, and connecting the mathematical + data science ideas can benefit students across the mathematical and data sciences. Instead of thinking of data science as just a set of tools or skills, we believe that data science students should be taught foundational ideas that underlie working with data; after all, the tools will probably be completely different twenty years from now but the important ideas will remain the same. This paper describes pedagogy familiar to mathematicians and statisticians which can be emphasized to support a strong data science foundation (and also benefit all students in the mathematical sciences, along the way).

Each of us is in a different place in our thoughts on connections between mathematics and data science. Some mathematicians have fully transitioned and consider themselves to be data scientists. Other mathematicians are working to understand what data science even \textit{is}. Which is to say that some instructors surely already incorporate ideas from data science in their mathematics courses (and vice versa). We hope to provide those instructors with new ideas for their classrooms. Other instructors may not have ever thought about connections between mathematics and data science. We hope to reassure those instructors that they should not fear embracing data science and that they can bring data science connections into their classrooms. Indeed, we believe that our manuscript can be worthwhile for all instructors, no matter their thoughts on the connections between mathematics and data science.

\begin{figure}
\begin{center}
\includegraphics[scale=0.5]{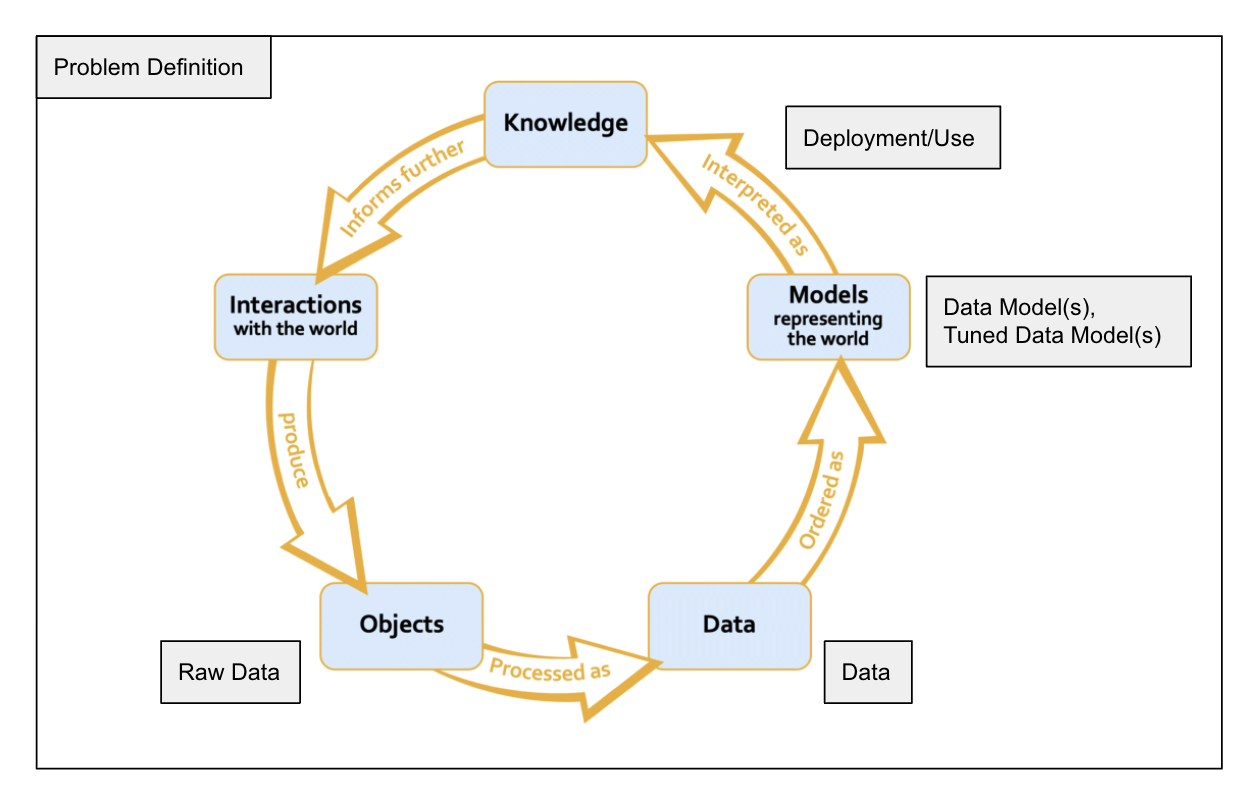}
    \caption{An example of the full data science lifecycle. \citep{colando2024}}
    \label{fig:ds-process}
\end{center}
\end{figure}

In 2018, The National Academies of Sciences, Engineering, and Medicine (NASEM) put out a report {\em Data Science for Undergraduates: Opportunities and Options} \citep{nas2018} outlining a vision for the emerging discipline of data science at the undergraduate level. A key goal of the report was to define what would give all students the ability to make good judgments, use tools responsibly and effectively, and ultimately make good decisions using data. The collection of those abilities are defined in the report as ``data acumen.''  Figure~\ref{fig:acumen} summarizes the components of data acumen\citep{nas2018}. Figure~\ref{fig:ds-process} provides an example of the data science lifecycle which is intimately connected to the items in Figure~\ref{fig:acumen}. We use the NASEM list to ground the discussion that follows, which hopes to connect data science curricula to the more familiar pedagogy used by many mathematical scientists. Whether you are a mathematician newly engaging with data science or a data scientist reflecting on its connections to mathematics, the argument that \textit{applying data science requires critical thinking}, something math instruction can nurture, is relevant to you. The {\bf {\em items in bold italics}} are ones that are considered in this article. 

\begin{figure*}
\begin{multicols}{2}
Mathematical foundations
\begin{itemize}
\setlength\itemsep{0em}
\item Set theory and basic logic
\item Multivariate thinking via functions and graphical displays
\item Basic probability theory and randomness
\item Matrices and basic linear algebra
\item Networks and graph theory
\item {\bf {\em Optimization}}
\end{itemize}

Computational foundations
\begin{itemize}
\setlength\itemsep{0em}
\item Basic abstractions
\item Algorithmic thinking
\item {\bf {\em Programming concepts}}
\item Data structures
\item Simulations
\end{itemize}

Statistical foundations
\begin{itemize}
\setlength\itemsep{0em}
\item Variability, uncertainty, sampling error, and inference
\item {\bf {\em Multivariate thinking}}
\item Nonsampling error, design, experiments (e.g., A/B testing), biases, confounding, and causal inference
\item Exploratory data analysis
\item Statistical modeling and model assessment
\item Simulations and experiments
\end{itemize}

Data management and curation
\begin{itemize}
\setlength\itemsep{0em}
\item Data provenance
\item {\bf {\em Data preparation, especially data cleansing and data transformation}}
\item Data management (of a variety of data types)
\item Record retention policies
\item Data subject privacy
\item Missing and conflicting data
\item Modern databases
\end{itemize}

\newpage
Data description and visualization
\begin{itemize}
\setlength\itemsep{0em}
\item Data consistency checking
\item Exploratory data analysis
\item {\bf {\em Grammar of graphics}}
\item Attractive and sound static and dynamic visualizations
\item Dashboards
\end{itemize}

Data modeling and assessment
\begin{itemize}
\setlength\itemsep{0em}
\item Machine learning
\item Multivariate modeling and supervised learning
\item Dimension reduction techniques and unsupervised learning
\item Deep learning
\item {\bf {\em Model assessment and sensitivity analysis}}
\item Model interpretation (particularly for black box models)
\end{itemize}

Workflow and reproducibility
\begin{itemize}
\setlength\itemsep{0em}
\item Workflows and workflow systems
\item Reproducible analysis
\item {\bf {\em Documentation and code standards}}
\item Source code (version) control systems
\item Collaboration
\end{itemize}

Communication and teamwork
\begin{itemize}
\setlength\itemsep{0em}
\item Ability to understand client needs
\item Clear and comprehensive reporting
\item Conflict resolution skills
\item {\bf {\em Well-structured technical writing without jargon}}
\item Effective presentation skills
\end{itemize}

Ethical problem solving
\begin{itemize}
\setlength\itemsep{0em}
\item {\bf {\em Ethical precepts for data science}} and codes of conduct
\item Privacy and confidentiality
\item Responsible conduct of research
\item Ability to identify ``junk" science
\item Ability to detect algorithmic bias
\end{itemize}
\end{multicols}
\caption{Key components of data acumen outlined by \citet{nas2018}. The bolded-italicized items are described in more details below. \label{fig:acumen}}
\end{figure*}

\section{Data science curricular foundations}

We will not endeavor to define data science (nor, for that matter, will we endeavor to define mathematics, statistics, or computer science).  Instead, borrowing from the {\em Curriculum Guidelines for Undergraduate Programs in Data Science}, data science can be described as an applied field with an emphasis on using data to describe the world, whose theoretical foundations are drawn primarily from the established disciplines of statistics, computer science, and mathematics \citep{parkcity2017}.

The established foundations, however, are not optimally effective without adjustments that connect to data driven analyses and decision making.  \citet{hardin2017} suggest that the needed mathematics content (to prepare students for data science) can be reconfigured into two courses: Mathematical Foundations I: Discrete Mathematics  (focused on linear algebra, counting, and graph theory) and Mathematical Foundations II: Continuous Mathematics  (focused on enough calculus to understand the ideas of partial derivatives (interactions in a model); approximating functions using Taylor series or Fourier series; probability as area/integration; multivariate thinking (functions, optimization, integration)) \citep{hardin2017,rad2024}. While we understand that revamping curricula is not a trivial task, we also recognize that reflecting on what is taught and why it is taught is a useful exercise. Certainly, the majority of students in our mathematics classes will not end up being data scientists (as of right now); however, the majority of our students {\bf will} end up engaging with data-driven ideas in their post-collegiate years.  It serves us well to understand what we are teaching and how we are teaching it with a lens toward helping our students navigate a world which is extraordinarily different than the world in which we were trained.

In what follows, we investigate some of the data acumen components (in {\bf {\em bold and italics}} in Figure~\ref{fig:acumen}) under two parallel lenses:  how is the idea important for data science?  how can mathematical sciences and mathematical pedagogy connect to the data science topic being presented? The subcomponents were chosen somewhat arbitrarily, although some of them are admittedly my favorite parts of teaching data science or the subcomponents that most closely align to mathematical thinking. I do not argue that every subcomponent is directly connected to a mathematical concept; mathematics and data science are, after all, different disciplines.

\subsection{Mathematical foundations: optimization}

We will not linger on mathematical foundations, as certainly you have already spent considerable time thinking about how your pedagogy can best communicate key mathematical concepts that are important for doing data science.

However, we highlight {\bf {\em optimization}} as a concept that may be so familiar to mathematicians that they don't think to emphasize its importance in, among other topics, machine learning.  We use derivatives as an amazingly powerful hammer used to identify optimal model parameters by minimizing or maximizing performance metrics, and students are able to easily parrot ``take a derivative" when asked how to maximize a function. But focusing on {\em why} the quantity should be optimized is often under-emphasized, leading the student to believe that the method is a black box. Many concepts from calculus are used in optimization, and explicitly connecting these concepts to problems in machine learning can be a powerful motivator for students interested in data science.  For example, the implementation of a gradient descent algorithm illustrates the use of directional derivatives, convergence of iterated functions, and multivariate probabilities to optimize complex loss functions, fit statistical models, and train machine learning algorithms efficiently and accurately. Note that the mathematics (e.g., gradient descent) is almost always hidden within the computational software, making it even more important that our students understand the underlying mathematics.

\subsection{Computational foundations: programming concepts}

There is evidence that the most common programming languages used in introduction to data science courses are R and Python, for example, see Table 1 in \citet{cetinkaya-rundel2021}. Ancillary skills such as regular expressions and SQL are key tools in a data scientists' toolbox. Certainly, you do your students a service if you provide them practice with common programming languages that they can take directly to a research project or to the workplace. 

But as educators, it is important to remember that we are {\em not} simply teaching a series of functions, or even a programming language. Twenty years from now, there may be new languages or new ways of approaching data driven decision making.  Instead of teaching rote skills, we should be communicating an approach to computing and ideas of {\bf {\em programming concepts}}, for example, iterations, recursion, vectorization.  Asking students why we are using the functions we use and to write out (in words!) what the computer program is doing will help students both learn the programming language and also move on to different computing approaches, when and if that adjustment is needed.

\citet{pruim2023} describe ``Better Coding Practices for Data Scientists."  There have been many previous calls for better coding practice, but \citet{pruim2023} describe the four Cs for good code: correctness, clarity, containment, and consistency.  They provide a scaffold for both the instructor and the student to understand not only what works but also how the code works within the context of doing data science well.  Developing new machine learning methods, extending existing ones, and comparing competing algorithms all requires strong computational foundations as outlined by \citet{pruim2023}.

As mathematical scientists, we liken good coding practices to good proof writing. It is vital that a proof is {\em correct}, but being correct is only the first step of the process.  The proof must also be {\em clear} so that, as \citet{pruim2023} state (modified to describe proofs instead of programs), ``humans reading and writing the [proof] can tell what it is intended to do, and easily make modifications as necessary."  The proof should be {\em contained} to include only the information needed to make the full argument.  And last, the proof should be internally {\em consistent} with notation and style.

As a mathematical scientist, you may not ever be tasked with teaching computational skills.  But you might be tasked with teaching an introduction to data science course.  In such a course, you will undoubtably encounter some programming assignments.  Approaching the teaching of coding as you would the teaching of proof writing will encourage your students to write better code and will benefit you to understand the importance of writing better code.

\subsection{Statistical foundations: multivariate thinking}

While much work has been done in recent years to modernize the introductory statistics course, the framework for the content of the course is well-established and an important part of the data scientist's toolbox (in particular, it focuses on the process of drawing conclusions about larger populations based on data, with an understanding of the variability and uncertainty inherent in the data, often referred to as statistical inference). That said, the majority of the content from introductory statistics does not necessarily belong in a foundations of data science course, it is important enough to warrant a full introductory statistics course as part of the data science curriculum. Indeed, the two courses should be distinct.

However, one topic which is fundamental to both statistics and also data science is {\bf {\em multivariate thinking}}.  Data science problems are inherently multivariable, and learning how to work with many variables simultaneously provides students the tools to understand a problem holistically.  

To introduce multivariate thinking, it is worth presenting examples where student intuition might be challenged enough for them to want to understand the mathematical underpinnings of the results. 
 Consider the idea of Simpson's Paradox where the directionality of the result is reversed when the effect is considered across subgroups.  Consider Table~\ref{tab:antibiotics} (taken directly from \citet{bonovas2023}) which describes a study to determine the feasibility of standardized surveillance of nosocomial infections (those infections that originate in the hospital) in patients (original study at \citet{severijen1997}, summarized in \citet{bonovas2023,norton2015}).  The relative risk (RR) measures the probability of urinary tract infection in the Yes (prophylactic antibiotics) versus the No (no prophylactic antibiotics) groups.  When broken down across the type of hospital (low-incidence hospitals are those where the UTI rate is less than 2.5\%), it seems clear that the use of prophylactic antibiotics is actually detrimental to patients.  When using the overall/global average, however, the information is misleading, prophylactic antibiotic use seems to be beneficial.  The paradox comes about because the proportion of patients who use prophylactic antibiotics varies so widely across the two types of hospitals. If the students want to understand the ideas more deeply, the global average can be worked out as a function of the individual hospital values.

 Of course, along with Simpson's paradox, there are many other multivariable effects to consider in many analyses: confounding, effect modification, multicollinearity, subgroup analysis, overfitting, and bias-variance trade-offs. We would do well to return to these ideas as often as possible so that our students are attuned to seeing them in their own analyses.

\begin{table}[h!]
\centering
\begin{tabular}{lccc}
\toprule
\textbf{Antibiotic Prophylaxis} & \textbf{Yes} & \textbf{No} & \textbf{RR} \\
\midrule
\textbf{Low-incidence hospitals} & 20/1113 (1.8\%) & 5/720 (0.7\%) & 2.59 \\
\textbf{High-incidence hospitals} & 22/166 (13.3\%) & 99/1520 (6.5\%) & 2.03 \\
\textbf{All hospitals (aggregate)} & 42/1279 (3.3\%) & 104/2240 (4.6\%) & 0.71 \\
\bottomrule
\end{tabular}
\caption{Example of Simpson's Paradox: Antibiotic Prophylaxis Results \label{tab:antibiotics}}
\end{table}

\subsection{Data management and curation: data preparation, especially data cleansing and data transformation}

In many courses, it is difficult to know where in the curriculum students should learn how to manage data. Should there be a course or a learning outcome devoted to {\bf {\em data preparation, especially data cleansing and data transformation}}?  We argue that data management (the practice of collecting, storing, organizing, protecting, and using data) is both a fundamental part of data science but additionally, when taught well, it is an important part of having an understanding of the data science problem. Data curation is a perfect example of teaching important tools while communicating that the tools are extremely likely to be very different in twenty years. Which is to say that the students {\bf must} learn the tools, but they must also learn the more important foundational ideas underlying the tools.

Consider the {\bf tidyverse} \citep{wickham2019} which is a dialect of R designed for solving data science challenges.  A full college-level course may not be needed to introduce basic data wrangling skills. And, indeed, the {\bf tidyverse} is fundamentally human-centered (e.g., careful thought has gone into naming of each function), which makes the functions straightforward for students to understand. However, there are important pedagogical reasons to include teaching data wrangling as a part of any (all?) data science courses.

We expect you agree that it is much more difficult for a student to grapple with complex ideas in Abstract Algebra if they do not have an understanding of basic algebra.  To that end, lower level competencies often inform our ability to perform higher level tasks.  While we do not argue that all data scientists must be computer science majors, we do contend that competency in working with data through coding/programming is vital to doing data science. Building on Wickham's {\bf tidyverse} \citep{wickham2019} package in \textbf{R}, \citet{erickson2019} describe data moves (equivalent to the data verbs in the {\bf tidyverse}, such as \texttt{mutate()}, \texttt{filter()}, \texttt{select()}, etc., which add columns, filter rows, and select columns of a data frame, respectively) and describe how they can and should be used in the function.  In their appeal to educators, they advocate to ``include data moves explicitly as a part of data analysis."  We advocate slightly more strongly: include data moves explicitly as a part of {\bf every} data analysis.

There are multiple ways to build up basic data management competencies. \citet{mcnamara2024} compares using the {\bf tidyverse} versus a different commonly used approach -- the formula syntax in {\bf R}.  Her in-class experiment demonstrates that there is not a single best approach to teaching programming (and data management).  She reiterates the importance of teaching only one syntax and being as consistent as possible with syntax \citep{mcnamara2021}. \citet{tidyeduc2022} describe the {\bf tidyverse} in more detail, focusing on the pedagogical benefits and opportunities over other tools for data management.

\subsection{Data description and visualization: grammar of graphics}

Bringing data visualization into the classroom allows for nuanced conversations about both how to do it well and why one should do it well.  If students are building their data viz skills using AI and message boards, their tool box will be a hodge-podge of techniques that make it difficult for them to build out their tools to use for the next task in front of them.

\begin{figure}[htbp]
  \centering
  \begin{subfigure}[b]{0.45\textwidth}
    \includegraphics[width=\textwidth]{images/cholera.jpg}
    \caption{John Snow's map of the 1854 cholera outbreak in London. Ask: \textit{why} is this graph effective? \citep{snow1855}}
    \label{fig:cholera}
  \end{subfigure}
  \hfill
  \begin{subfigure}[b]{0.45\textwidth}
    \rotatebox{1}{\includegraphics[width=\textwidth]{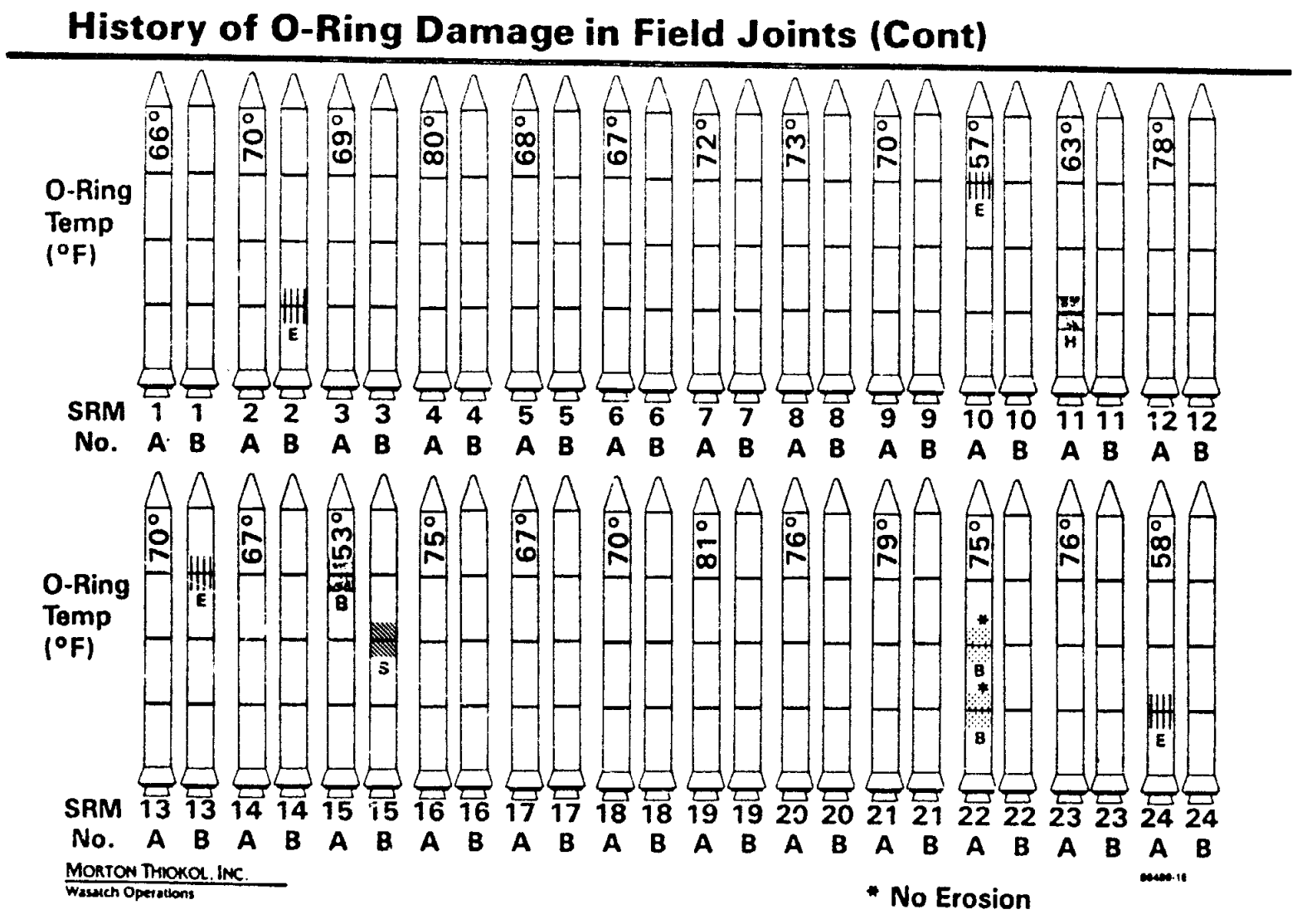}}
    \caption{Images used to convey information about data collected before the Challenger disaster. Ask: \textit{why} is this graph ineffective? \cite{rogers1986}}
    \label{fig:challenger}
  \end{subfigure}
  \caption{Images that allow students to think carefully about what aspects of a graph make it effective or ineffective.}
  \label{fig:sidebyside}
\end{figure}

Edward Tufte's, ``Visual and Statistical Thinking: Displays of Evidence for Making Decisions" \citep{tufte1997} is an excellent booklet for teaching data visualization.  Tufte describes two real settings where visualizations played an important role in the decision making process (John Snow's work in the London cholera epidemic of 1854 and the Challenger explosion in 1986, see Figure~\ref{fig:sidebyside}).  

\begin{figure}
    \includegraphics[scale = 0.35]{"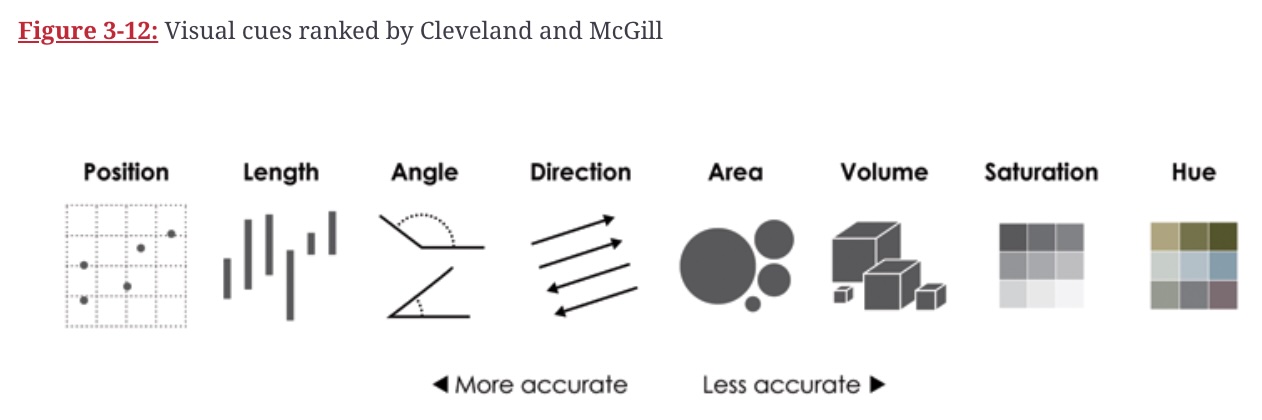"}
    \caption{Image 3-12 in \citet{yau2013}, based on work by \citet{cleveland1984} \label{yau}}
\end{figure}

An alternative to asking what is ‘good’ or ‘bad’ about a particular graph is to use data visualization research to help formulate which aspects of the graph are effective at conveying the relevant message. For example, we use \citet{cleveland1984} to describe how some visualization types (e.g., scatterplots which use \textit{position}) are more accurate than other visualization types (e.g., heatmaps which use \textit{hue}), see Figure~\ref{yau}.  \citet{yau2013} describes visual cues which are the foundational pieces of the {\bf {\em grammar of graphics}} (which is described in detail by \citet{wilkinson2005} and led to the framework of the {\bf ggplot2} package\citep{wickham2016}). We break down each graph into its constituent parts and describe which parts are needed, which parts are spot-on, and which parts can be improved.  The discussion leads directly into a conversation about data visualization and the tools needed to build data visualizations that can convey the desired message.

As previously mentioned, the exact tools (e.g., the {\bf tidyverse}, {\bf ggplot2}, or even {\bf R}) that we use are not the point. The point, instead, is to teach students how to break down a problem into logical parts (wrangling) and to put together a visualization that includes best practices for visual displays \citep{few2012}. In the cholera and Challenger examples (see Figure~\ref{fig:sidebyside}), we follow \citet{nolan2016} to ask: Does the image \textit{make the data stand out}, \textit{facilitate comparisons}, and \textit{add information}? It is only after really understanding the lower-level tools (not just how to employ them but {\bf when} and {\bf why} to employ them) that students will be able to perform higher level data science skills.  A good exploratory data analysis that includes effective data visualization informs model building and subsequent inferences.
 
\subsection{Data modeling and assessment: model assessment and sensitivity analysis}

Simulation brings together ideas of iteration with ideas of {\bf {\em model assessment and sensitivity analysis}}.  \citet{morris2019} describe the importance of simulations to evaluate statistical models.  Additionally, they convey best practices to help students formulate simulation studies which are optimally suited for their research question. \citet{loy2021} describes using simulations to do statistical inference on sets of permuted images.

Simulations can also be used to understand ethical implications of models applied to heterogeneous populations.  A compelling (made-up) example by Aaron Roth describes a single model used to accept students into college from two groups: one group who took the SAT one time, one group who took the SAT twice and submitted the higher of the two scores \citep{roth2019}.\footnote{I present the full simulation in my class notes at \href{https://st47s.com/Math154/Notes/04-simulating.html\#biasmodels}{https://st47s.com/Math154/Notes/04-simulating.html\#biasmodels}.}  The simulation study reveals the following about applying one model to two populations:
\begin{itemize}
\item 
Depending on the nature of the difference in the two groups, a single model can be either to the benefit or the detriment of the minority population.
\item 
The problem comes with the variable {\ttfamily SAT} meaning different things across the two groups. There was no explicit human bias, either on the part of the algorithm designer or the data gathering process.
\item 
The problem is exacerbated if we artificially force the algorithm to be group blind, that is, if we are forced to use a single model on the variable {\ttfamily SAT}, even if it means different things across the two groups.
\item 
Well-intentioned ``fairness" regulations prohibiting decision makers from taking sensitive attributes into account can actually make things less fair and less accurate at the same time.
\end{itemize}

By harnessing ideas of iteration and simulation, both of which are important topics in many mathematical courses (and are components of data acumen in both the Computational foundations and Statistical foundations, see Figure~\ref{fig:acumen}), we are able to reflect on what we know and don't know about a particular model of interest. Note that simulations are also key aspects of the data lifecycle (see Figure~\ref{fig:ds-process}), important in tuning models as well as in understanding the outcome of models deployed in the world.

\subsection{Workflow and reproducibility: documentation and code standards}

In mathematics, notation matters.  Indeed, notation is fundamental to communicating extremely abstract and complicated mathematical concepts. Even simple algebraic equations would be difficult to describe without an agreed upon language of polynomials, arithmetic, numbers, and variables. While your own notation may lead you to correct mathematical discoveries, it will be very difficult to collaborate or to communicate your ideas if others do not share an understanding of your notation.

The same is true in computing.  Notation matters.  Consistent {\bf {\em documentation and code standards}} are key to being able to communicate ideas through computing.  Without an agreed upon standard for writing code, collaboration and communication becomes difficult. Different languages use different styles, and for students in the early stages of learning, it is important to communicate using a single style and sticking to it.  If you are teaching with the {\bf tidyverse}, you can introduce your students to the {\em Tidyverse Style Guide} \citep{style}. If you are teaching with a different dialect or programming language, you can introduce your students to the style guide associated with the coding choices you are making. I personally teach with the \textbf{tidyverse}, primarily because it eases the learning process and cognitive load for users when compared with other dialects \citep{tidyeduc2022}.

\subsection{Communication and teamwork: well-structured technical writing without jargon}

One of the hardest parts of communicating a data science project is knowing how much information (e.g., code) to provide and how much to leave out.  Creating {\bf {\em well-structured technical writing without jargon}} takes not only practice but also thoughtful consideration of getting ideas across. Mathematics is not done in a vacuum, and neither is data science. Mathematicians need to successfully communicate results, and the impact of those results, to broad, diverse audiences.  In academia this is how we get funding, publish papers, and get invited to give talks.  In industry and government it is how we come together to find a solution to a problem and to disseminate the impact of our proposed solution.

The literature is full of calls to engage students with writing in order to deepen their understanding of mathematical topics.  \citet{woodard2020} write about how not only does writing help students but that the writing assignments provide a ``unique approach toward assessing the students' understanding of statistical concepts.''  \citet{quealy2014} describes how ``writing can be a vital instrument in the learning process" of mathematical ideas. \citet{drucker2018} focus on the writer understanding the needs, goals, and knowledge of the intended audience.  

While recognizing the challenges that writing assignments pose (e.g., challenges in scaling to large classes, potential for cheating with ChatGPT, difficulty with consistent grading, etc.), figuring out ways to encourage writing in the classroom communicates the value of writing to our students. We do them a disservice if they graduate having never written a paper or report in mathematics, statistics, or data science. Learning how to communicate about technical ideas transcends the specific disciplinary content and encourages students to think beyond themselves to diverse audiences: the public, policy-makers, scientists, and fellow mathematicians / statisticians / data scientists.

\subsection{Ethical problem solving: ethical precepts for data science}

While seemingly outside of the mathematician's toolbox, connecting data science conclusions to an ethical framework is important to convey to students.  Philosophers have been debating ethical questions for millennia, including the Chinese philosopher Confucius (551–479 BCE) and the Greek philosopher Socrates (469–399 BCE). Bringing up connections between data science quandaries and ethical approaches to decision making teaches students that their work lives in a much broader context than just the classroom \citep{colando2024}. One does not need to be a philosopher or even a data scientist to find examples which dig into the {\bf {\em ethical precepts for data science}}.  For example, many ideas can be generated by exploring the resources at the Markkula Center for Applied Ethics \href{https://www.scu.edu/ethics/}{https://www.scu.edu/ethics/}.

\section{Discussion}

The nascent but flourishing discipline of data science is grounded in centuries of mathematics and statistics. Data science builds on the technical results and practices from mathematics and statistics.  But more importantly, data science builds on an approach to quantitative thinking that facilitates novel discovery and progress. As educators, it is in our best interest to teach using best practices for building strong data science students. Those best practices will benefit not only students going on to do data science, but also other students: from those who will not pursue quantitative fields but will become data consumers all the way to those who will obtain PhDs in theoretical mathematics. Working with the foundational concepts in data science is crucial to building a mathematical sciences curriculum that will serve tomorrow's leaders.

\section*{Acknowledgments}

Thank you to the reviewers for their valuable suggestions. Thank you to Nicholas J. Horton and Rich Levine for ideas and discussion.

\newpage

\bibliographystyle{apalike}

\bibliography{ds_ams.bib}

\end{document}